\documentclass[a4paper,conference]{IEEEtran}
\usepackage{color}
\usepackage{array}
\usepackage{verbatim}
\usepackage{booktabs}
\usepackage{graphicx}
\usepackage{subcaption}

\makeatletter

\def\ps@IEEEtitlepagestyle{%
  \def\@oddfoot{\mycopyrightnotice}%
  \def\@evenfoot{}%
}

\newcommand\copyrighttext{%
  \footnotesize \textcopyright~2017 IEEE. Personal use of this material is permitted. Permission from IEEE must be obtained for all other uses, in any current or future media, including reprinting/republishing this material for advertising or promotional purposes, creating new collective works, for resale or redistribution to servers or lists, or reuse of any copyrighted component of this work in other works.}

\def\mycopyrightnotice{%
  {\fbox{\parbox{\dimexpr\textwidth-\fboxsep-\fboxrule\relax}{\copyrighttext}}}
  \gdef\mycopyrightnotice{}
}

\usepackage[factor=1100,stretch=10,shrink=20,tracking=true,kerning=true,spacing=true]{microtype}

\usepackage{textcomp}

\usepackage {url}
\usepackage [numbers]{natbib}

\def\baselinestretch{0.95}
\usepackage[font=small]{caption}
\usepackage{paralist}

\makeatother

\begin{document}

\title{SDN Enhanced Ethernet VPN for Data Center Interconnect}

\author{\IEEEauthorblockN{Kyoomars Alizadeh Noghani, Andreas Kassler}
\\
\IEEEauthorblockA{Karlstad University
\vspace{-2ex}
\\\\
\{kyoomars.noghani-alizadeh, andreas.kassler\}@kau.se}}

\maketitle

\begin{abstract}
Ethernet Virtual Private Network (EVPN) is an emerging technology that addresses the networking challenges presented by geo-distributed Data Centers (DCs). One of the major advantages of EVPN over legacy layer 2 VPN solutions is providing All-Active (A-A) mode of operation so that the traffic can truly be multi-homed on Provider Edge (PE) routers. However, A-A mode of operation introduces new challenges. In the case where the Customer Edge (CE) router is multi-homed to one or more PE routers, it is necessary that only one of the PE routers should forward Broadcast, Unknown unicast, and Multicast (BUM) traffic into the DC. The PE router that assumes the primary role for forwarding BUM traffic to the CE device is called the Designated Forwarder (DF). The proposed solution to select the DF in the EVPN standard is based on a distributed algorithm which has a number of drawbacks such as unfairness and intermittent behavior. In this paper, we introduce a Software-Defined Networking (SDN) based architecture for EVPN support, where the SDN controller interacts with EVPN control plane. We demonstrate how our solution mitigates existing problems for DF selection which leads to improved EVPN performance.
\end{abstract}

\begin{IEEEkeywords}
Cloud Networking, Ethernet Virtual Private Network, EVPN, Software Defined Networks, SDN, Designated Forwarder.
\end{IEEEkeywords}


\section{Introduction}
\label{sec:intro}
Ethernet Virtual Private Network (EVPN)~\cite{rfc7432} has been recently proposed to provide a flexible and scalable Layer 2 (L2) interconnection among geo-distributed Data Centers (DCs) and tenants. EVPN distributes MAC address reachability information in control plane using MP-BGP protocol and mitigates legacy L2VPN scalability problems. Providing All-Active (also known as Active-Active) (A-A) mode of operation is another advantage of EVPN over preceding L2VPN technologies so the traffic can truly be Multi-Homed (MH) on Provider Edge (PE) devices.

\begin{figure}[tbh]
\begin{centering}
\includegraphics[width=0.47\textwidth]{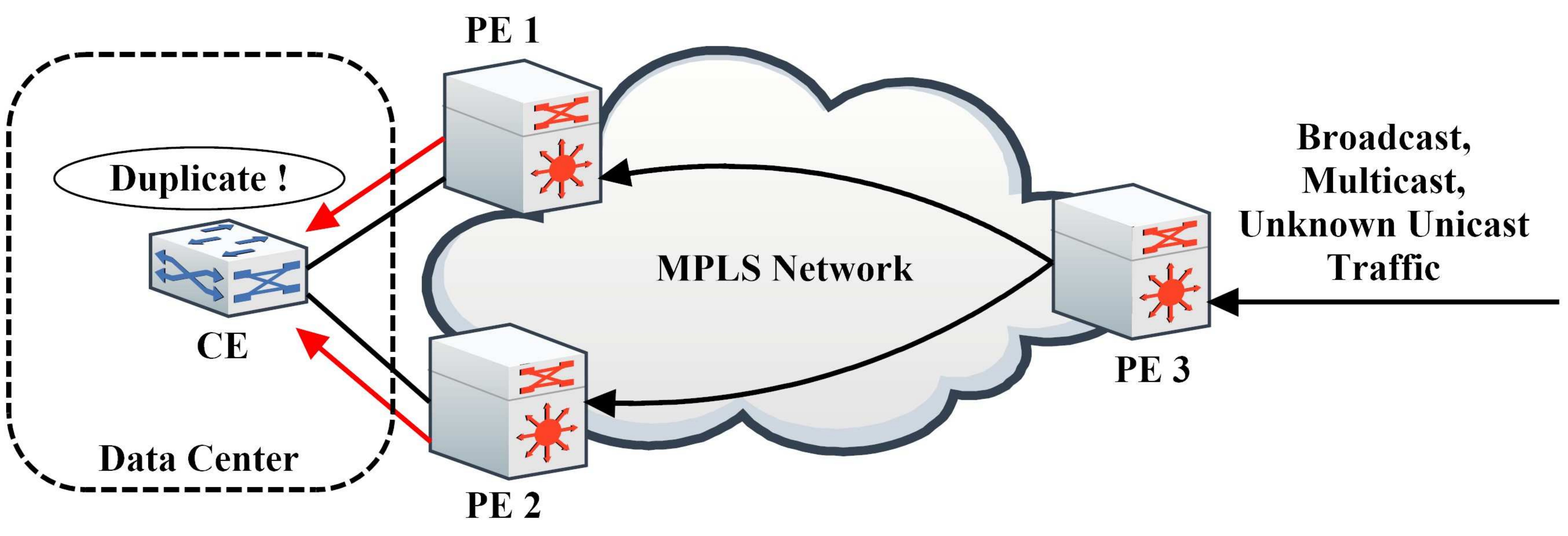}
\par\end{centering}
\caption{\label{fig:DF}Routing challenge for BUM packets}
\vspace{-2mm}
\end{figure}

Although A-A mode of operation is a very beneficial feature, it introduces new challenges such as importing multi-destination traffic (Broadcast, Unknown unicast\footnote{Traffic for which a PE does not know the destination MAC address.}, and Multicast (BUM)) multiple times into the DC (see Figure~\ref{fig:DF}). Importing BUM packets through multiple routers into the DC leads to an undesirable flooding, overhead, and disruption. As a result, PE routers participating in the same EVPN Instance (EVI\footnote{EVPN Instance (EVI) identifies an Ethernet VPN in the MPLS network.}) must agree among themselves as who should act as the Designated Forwarder (DF). The DF is responsible for forwarding BUM traffic on a particular Ethernet Tag\footnote{An Ethernet tag identifies a particular broadcast domain, such as a VLAN.} and Ethernet Segment (ES)\footnote{A set of Ethernet links that connect a MH device to a BGP router.} to the Customer Edge (CE) router.

The default DF election algorithm defined by EVPN standard~\cite{rfc7432} is called ``service-carving'' which is a distributed algorithm that each PE runs independently in order to perform load balancing for multi-destination traffic destined to a given ES. However, service-carving encounters a number of fundamental problems such as inconsistent output, undesirable DF swap, and fairness problems. Although a number of distributed solutions have been proposed~\cite{hao-bess-evpn-df-handshaking-02, ietf-bess-evpn-df-election-02, rabadan-bess-evpn-pref-df-02, sajassi-bess-evpn-fast-df-recovery-00} to improve the service-carving algorithm, neither of them could fully address the problems.

Our previous work~\cite{7939142} proposes a Software-Defined Networking (SDN)~\cite{onf} based framework that automates the EVPN management in geo-distributed DCs. In this paper, we extend the proposed framework to improve the functionality of EVPN by selecting the DF from the SDN controller in a more intelligent way. In contrast to a distributed solution, which only considers the physical existence of PE, in SDN-based solution, the controller centrally instructs the PEs to switch-over the DF role according to different criteria, such as port or link utilization. Furthermore, the SDN-based solution helps the administrator to select the desired DF by sending high-level commands without being involved in the complexity of configuring PEs. The service-carving algorithm as well as a handshake method~\cite{hao-bess-evpn-df-handshaking-02} that is developed to cover the service-carving problems are compared with the SDN-based approach. We show how the comprehensive view over the network using the SDN architecture helps to select an appropriate DF leading to lower overhead and better performance.

The remainder of this paper is organized as follows. Section II presents the relevant background for our work. Section III describes the proposed architecture. In Section IV, we describe and present the results of our experiments. Finally, this paper is concluded in Section V.


\section{Background}
\label{sec:background}
The service-carving algorithm, defined in EVPN RFC~\cite{rfc7432}, works as follows:

\begin{enumerate}
\item When a PE discovers the ES Identifier (ESI) of the attached ES, it triggers a specific MP-BGP message (ES route~\cite{rfc7432}).

\item The PE then starts a timer to allow the reception of ES routes by other PE nodes connected to the same ES.

\item The receiver PEs also start a timer when the ES route is received. This timer value should be the same across all PEs connected to the same ES. 

\item When the timer expires, each PE starts DF election process independently using the same algorithm. The default DF election algorithm is based on a (V mod N) = I function that provides a local DF election of a PE at $<$ESI, EVI$>$ level for a given ES. V is the Ethernet Tag associated to the EVI and N is the number of PEs for which ES routes have been successfully imported.

\item The PE elected as DF for a given EVPN instance unblocks the BUM traffic in the egress direction towards the DC while the non-DF PEs block the traffic immediately. 
\end{enumerate}

There are various concerns regarding the aforementioned way of selecting the DF. Firstly, dual DF may coexist during the DF re-election transient period since each PE relies on an independent timer to trigger the local DF election process. This problem leads to transient routing loops, flooding, and disruption in the network. Secondly, any change in the physical status of PEs (such as boot-up, failure, or recovery) triggers DF re-election procedure for all VLANs which consequently may lead to undesirable DF swap and causes service interruption. Thirdly, the algorithm does not perform fair when the Ethernet Tag follows a non-uniform distribution, for instance when the Ethernet Tags are all even or all odd. Likewise, using the service-carving algorithm, it may happen that one of the PEs does not get elected as the DF, so it does not participate in the DF responsibilities at all. Fourthly, the proposed solution does not take into account the network condition (such as links or ports utilization) in selecting the DF. Finally, the network operator is not able to choose the DF in a deterministic way.

To address the aforementioned problems, multiple RFC drafts have been proposed to improve the DF election procedure.  Hao et al.~\cite{hao-bess-evpn-df-handshaking-02} proposed handshake mechanism to avoid packet duplication by providing a better coordination among PEs. Mohanty et al.~\cite{ietf-bess-evpn-df-election-02} address the fairness problem of the service-carving algorithm and propose a new hash-based function according to Highest Random Weight (HRW). Rabadan et al.~\cite{rabadan-bess-evpn-pref-df-02} leverage the HRW and handshaking mechanism and further try to address other problems such as synchronization among PEs. Moreover, authors propose to make the DF re-election procedure more deterministic and not be influenced out of control by changes in PE status. Sajassi et al.~\cite{sajassi-bess-evpn-fast-df-recovery-00} propose how to improve the service-carving recovery procedure upon link or node failure. Although the proposed solutions mitigate a number of problems, they are not dynamic, require manual configuration and need an extra message passing among PEs. Moreover, all proposed solutions are distributed and problems such as switch-over the DF role according to network conditions would never be addressed unless there is a holistic view of the network.

We propose to select the DF from a centralized control plane through an SDN controller, which solves synchronization and fairness problems and prevents unnecessary DF churn when PE routers boot-up or go down. In addition, by inspecting monitoring data by e.g., polling switch or flow counters, the SDN controller is able to change the DF dynamically according to the network status. Although Hao et al.~\cite{DF_centralized} have already proposed a centralized DF election method, it has never been deployed and evaluated. In this paper, we develop an SDN-based architecture for EVPN management in DCs, investigate how the SDN controller can select the DF based on the status of the multicast tree, and evaluate its performance implications.

\section{Proposed Architecture}
\label{sec:architecture}
\begin{figure}[tbh]
\begin{centering}
\includegraphics[width=0.47\textwidth]{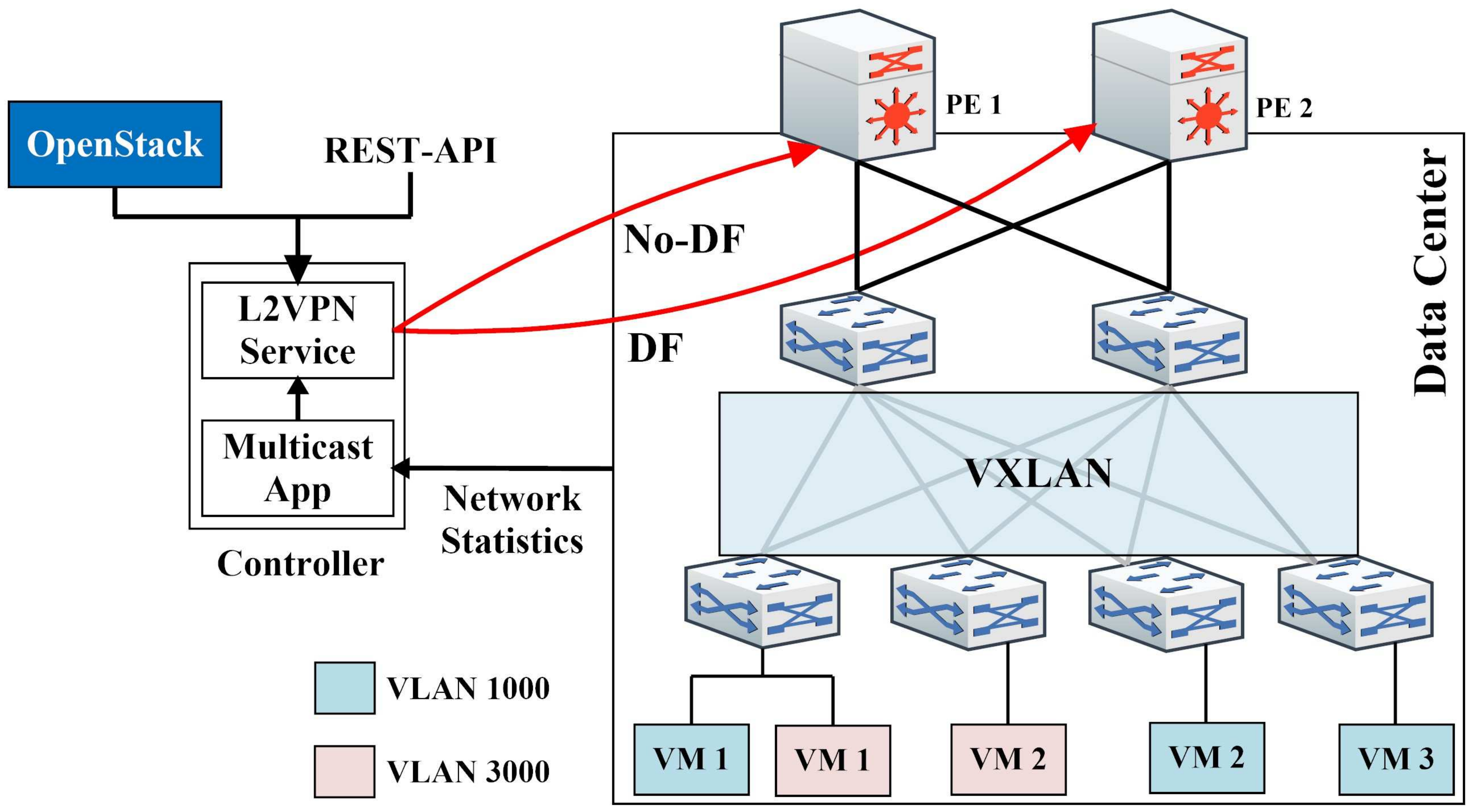}
\par\end{centering}
\caption{\label{fig:DF-SDN}Proposed SDN-based architecture for EVPN management}
\vspace{-2mm}
\end{figure}

In this paper, we assume that an SDN controller is managing the whole DC network based on its topological view. Another key responsibility of the controller is to select the appropriate DF for each ES and change it when needed. Herein, we assume that the DC and the core network belong to the same network provider so the DC administrator is capable of managing all underlying network entities including PE routers. The controller is directly connected to PEs using out-of-band control channel. Therefore, the controller communication with PEs is not affected by data plane congestion. The architecture is depicted in Figure~\ref{fig:DF-SDN}.

\vspace{-3mm}

The controller has the following two modules: (1) Multicast application and (2) L2VPN Service. The multicast module is in charge of establishing and managing multicast trees for each logical network by installing proper OpenFlow forwarding rules. The L2VPN Service is originally developed to automate EVPN deployment in a DC~\cite{7939142} and it knows all the PEs and EVPN parameters by implementing the required MP-BGP extensions and establishing MP-BGP connectivity with PEs. Moreover, the L2VPN Service helps the administrator to easily manipulate EVPN configuration on PE routers using NETCONF~\cite{rfc6241} protocol. In this paper, we extended the L2VPN Service module to select the DF for each EVI.

The SDN controller provides two ways for selecting the DF, deterministic and dynamic. In the deterministic approach, the network administrator (or a control application) selects the DF by sending high-level commands through available northbound interfaces such as Rest-API or OpenStack. Subsequently, the controller configures the current DF to block the BUM traffic and configures the desired DF to steer the traffic. In the dynamic mode, the controller chooses the appropriate DF according to different criteria such as PE utilization, link-level congestion, multicast tree status, network policy, etc. The controller then dynamically selects a new PE to take over the DF role that meets the requirements better. In this paper, we explore how the controller selects the DF dynamically according to the multicast tree status.

\subsection{BUM Traffic Routing}
Although this paper is mainly focused on selecting the DF, we ought to provide a background on how the BUM traffic belonging to a given EVPN is forwarded to the destination within the DC. This is particularly important given that the DC has to transport BUM traffic coming from an MPLS provider network and therefore, the interaction between inside the DC and the MPLS provider network must be clearly defined.

Network virtualization has become a popular topic in recent years that attempts to address the multi-tenant DC demands and network scalability aspects. Network virtualization can be achieved using techniques like overlay network technologies. Overlay networks are created by encapsulating and tunneling the traffic over a physical network. VXLAN~\cite{rfc7348}, NVGRE~\cite{rfc7637} or its variations such as MPLS over GRE are examples of network overlay technologies. Since VXLAN is one of the most widely used overlay technologies, we also assume that DC networks are VXLAN-based. VXLAN provides L2 extension over a shared L3 underlay by encapsulating Ethernet frames into IP User Data Protocol (UDP) headers and transports the encapsulated packets to the remote VXLAN Tunnel Endpoints (VTEPs) using standard IP routing and forwarding. There are two types of VTEPs: (1) virtual VTEPs are software-based and reside on a hypervisor in the servers and (2) hardware-based VTEP (typically a Top of Rack (TOR) switch). Herein, we assume that VTEPs are the latter type.

\begin{figure}[tbh]
\begin{centering}
\includegraphics[width=0.47\textwidth]{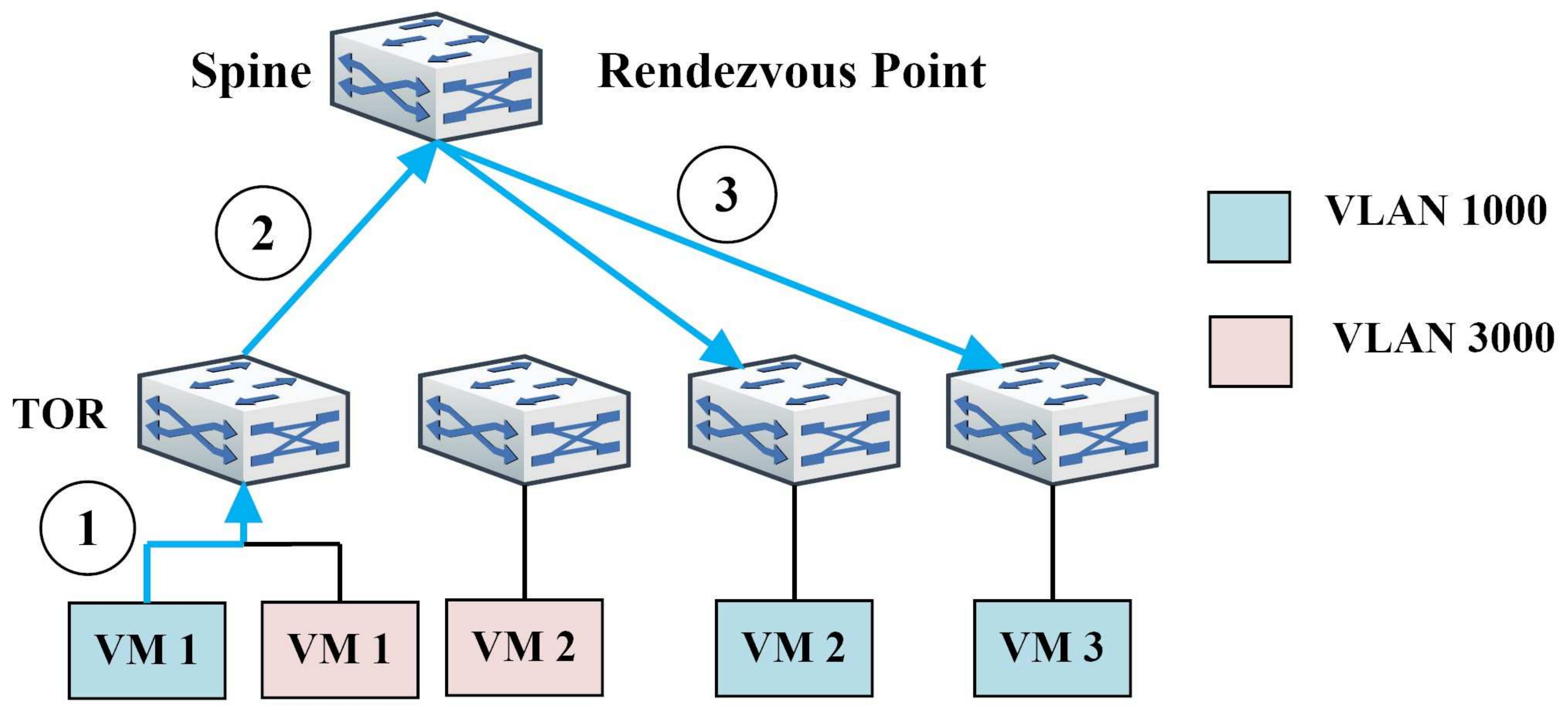}
\par\end{centering}
\caption{\label{fig:multicast}BUM traffic routing in a VXLAN overlay network}
\vspace{-3mm}
\end{figure}

The VXLAN translates broadcast messages in the virtualized subnet into multicast messages in the physical network. When a VM sends a BUM traffic (e.g., ARP), the VTEP has mainly two options to send that packet to the destination VTEPs: (1) Ingress replication~\cite{rfc7988} or (2) underlay IP multicast protocol such as PIM. In Ingress replication technique, whenever a VTEP must broadcast a frame (e.g., ARP packet) into a VXLAN segment, it replicates the frame in hardware and unicasts the frame to destination VTEPs. In contrast, in IP multicast protocol all VTEPs which belong to the same VXLAN Network Identifier (VNI) join the multicast tree. When a VTEP receives the broadcast frame, it encapsulates the frame in a VXLAN header and multicasts the encapsulated frame to the multicast address that is assigned to the VNI at the time of creation. The aforementioned procedure is demonstrated in Figure~\ref{fig:multicast}.

In Figure~\ref{fig:multicast}, VM-1 (which belongs to VLAN 1000) generates a broadcast frame (e.g., ARP). VTEP on TOR-1 encapsulates this broadcast frame into a UDP packet destined to multicast group address that is assigned to VXLAN 1000\footnote{For simplicity we assume that VTEP maps a VLAN to VXLAN with the same ID.} at the time of creation. The physical network delivers the packet to the Rendezvous Point (RP), which forwards it to other leaves that have already joined the multicast tree. TOR-3 and TOR-4 receive the packet while TOR-2 does not receive it since TOR-2 does not have any VM belongs to VLAN 1000. VTEPs on TOR-3 and TOR-4 remove the encapsulation header and deliver the packet to the corresponding VMs.

The same scenario happens for BUM traffic entering the DC. When a PE receives a BUM packet from the MPLS network, it maps the traffic to an appropriate VXLAN tunnel, sends the packet to the RP which subsequently transfers the packet to corresponding VTEPs inside the DC.

\subsection{Multicast Tree Inside a DC}
Using multicast tree may significantly reduce undesired traffic (such as BUM) within the DC. However, network providers are usually reluctant to deploy IP multicast due to concerns about security, reliability, and scalability, not to mention the requirement to have all routers in the network support the related protocols and be appropriately configured. Moreover, there are some other concerns about IP multicast deployment in DC networks. For instance, IP multicast is not designed to benefit from path diversity of DC networks which may result in poor bandwidth utilization. 

\begin{figure}[tbh]
\begin{centering}
\includegraphics[width=0.45\textwidth]{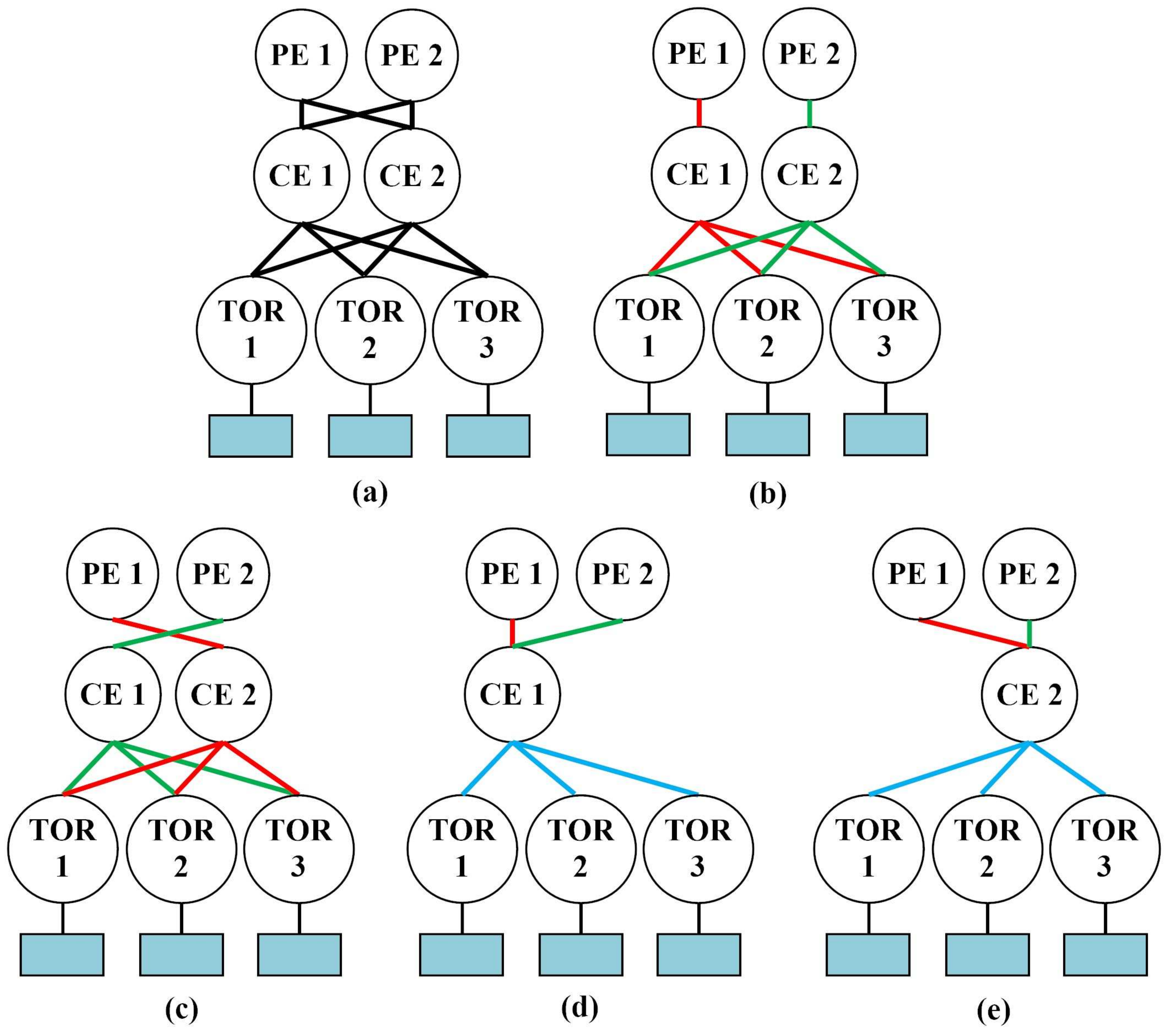}
\par\end{centering}
\caption{\label{fig:multicast-trees}Possible multicast trees within the DC}
\vspace{-2mm}
\end{figure}

Introducing centralized control into the multicast routing problem in DCs significantly improves the performance. Multicast routing algorithms can thus leverage topology information to build optimal routing trees and leverage link utilization state to efficiently exploit path diversity in DCs. To this end, a number of SDN-based solutions have been proposed in the literature (such as~\cite{6734903}) to deploy IP multicast in DC networks. Thanks to path diversity available in the DC network, there would be multiple multicast trees connecting TOR to PEs where EVPN BUM traffic can be routed over. The total number of multicast trees within the DC varies depends on the number of spine nodes, TOR nodes that participate in a given VXLAN network, PE routers participating in an EVI, and the number of available paths between them. Figure~\ref{fig:multicast-trees} demonstrates a sample network topology (Figure~\ref{fig:multicast-trees}.a) and possible multicast trees (Figure~\ref{fig:multicast-trees}.b-4.e) that the controller can select. The green and red lines show the multicast tree for PE-1 and PE-2, respectively, and the blue lines show the common links between multicast trees.

\subsection{Proposed Solution}
In our proposal, the multicast application inside the SDN controller establishes multicast trees where switches at the spine layer are RPs. Then, the multicast application continuously monitors multicast tree weights by polling switch counters and analyzing traffic passing through the links. Finally, the multicast application selects the tree with minimum total link weight as the active multicast tree for each VXLAN network. Accordingly, the PE router of the active multicast tree is the DF. By changing the active multicast tree, the SDN controller also needs to change the DF in order to send BUM traffic over a tree which has less probability of dropping a packet. In the evaluation section, we demonstrate how SDN approach could save BUM packet from being dropped and subsequently could prevent undesired BUM traffic to be re-sent.


To understand the procedure of changing the DF according to the multicast tree weight consider Figure~\ref{fig:multicast-trees}.b and assume that the multicast tree that is shown in red is selected by the controller as the active multicast tree for a specific VLAN (e.g., VLAN 1000). As a result, PE-1 would be the DF for this logical network. When the total traffic passes through the red links increase, the controller may find the green multicast tree a better option to transfer BUM traffic. Hence, the multicast application installs the corresponding rules on TOR switches to send BUM traffic to the CE-2 switch and informs the L2VPN Service module about the desired DF. Subsequently, the L2VPN Service asks the PE-1 router to drop the BUM packet belonging to VNI 1000 and configures PE-2 to transfer the BUM traffic towards the link to the RP (CE-2).

To avoid changing the active multicast tree too frequently, the multicast application may also need to consider other parameters. For instance, the multicast application may change the multicast tree, if its current total weight is not the minimum value and congestion level on one or some of its links are higher than a threshold for more than a specific time interval. This approach is considered in this paper.

\subsection{Using SDN Controller for DF Selection}
Based on the types of underlying network and the PE routers, the SDN controller has multiple options to select a PE to be a DF. In a DC, where the controller could interact with PE routers through a southbound interface such as OpenFlow protocol, the controller directly instructs the PE routers to block or unblock the BUM traffic belonging to the specific VLAN towards the DC. In DC networks where the PE router does not support OpenFlow protocol, the SDN controller may select the DF by triggering appropriate MP-BGP message (as proposed in~\cite{DF_centralized}) or use protocols such as NETCONF~\cite{rfc6241} to configure the PE to act as the DF.

\section{Evaluation}
\label{sec:eval}
In this section, we evaluate and compare the performance of the SDN-based solution for selecting the DF with alternative distributed solutions. First, we introduce the experimental methodology followed by an evaluation and discussion of our results.

\subsection{Experimental Methodology}
\begin{figure}[tbh]
\begin{centering}
\includegraphics[width=0.45\textwidth]{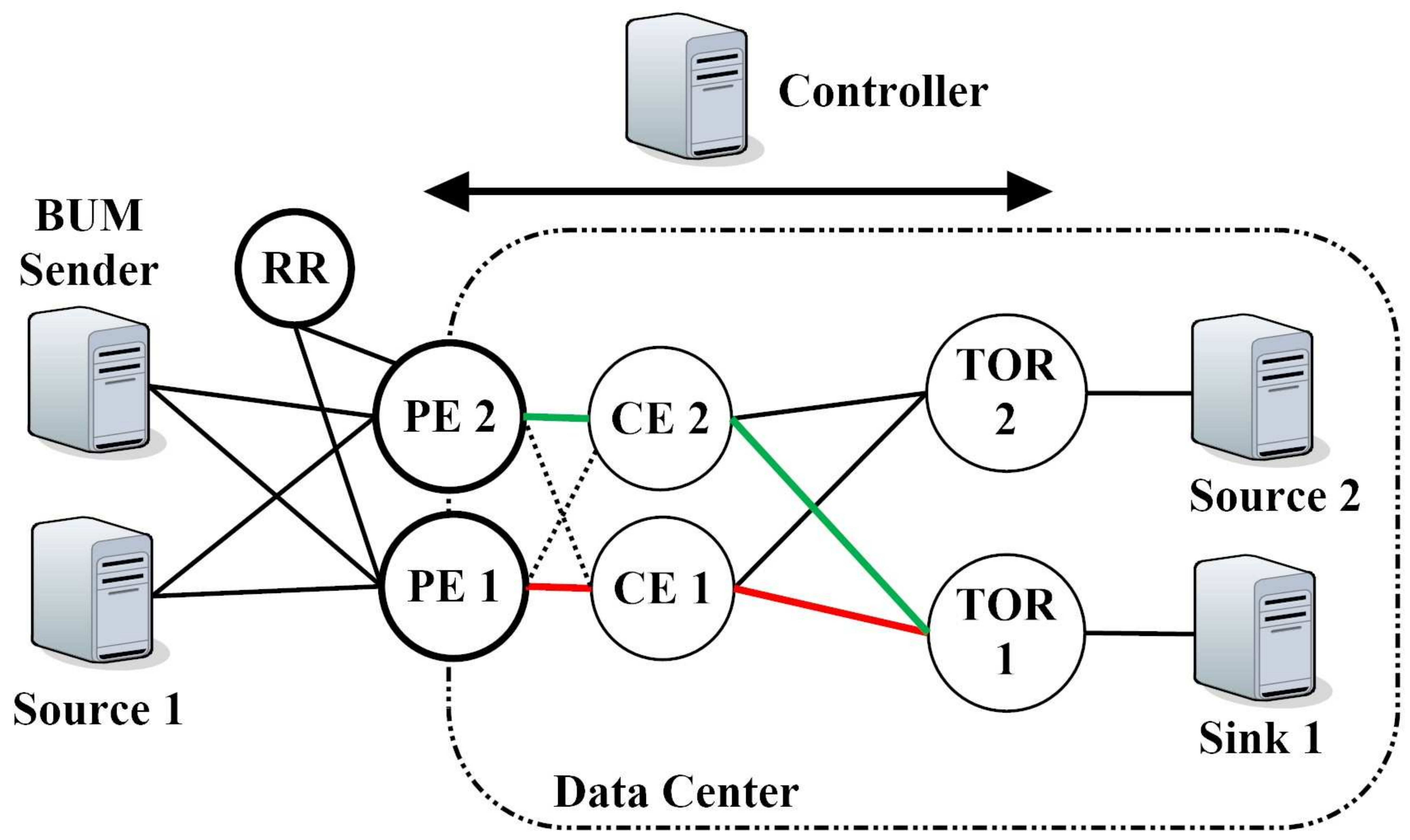}
\par\end{centering}
\caption{\label{fig:testbed}Experiment scenario}
\vspace{-2mm}
\end{figure}

Figure~\ref{fig:testbed} shows the network topology which is emulated using Mininet~\cite{mininet}. All switches in the topology are OpenFlow capable (vSwitch 2.6~\cite{ovs}), which are connected to the OpenDaylight SDN controller~\cite{Opendaylight}. The controller manages the DF as well as non-DF PEs by installing appropriate flow rules on them through OpenFlow protocol. All links are configured for 1 Gbps bandwidth and 1 ms delay. PE routers are interacting with each other through the Route Reflector (RR) located in the core network. Two senders (Source-1 and Source-2) generate background traffic (TCP) and one sender (BUM Source) generates broadcast packets (UDP) for the configured EVPN using Iperf. 

One client (Sink-1) is considered inside the DC topology to participate in the EVPN. The dashed links between PE and CE routers are needed to emulate an MH device but we consider them as fully congested so that the multicast application never selects the tree which contains one of these links. The red and green lines represent two possible multicast trees the multicast application may select. The red tree depicts the active multicast tree at the beginning of the experiment. The experiments run over a 3.2 GHz Core i7 processor Intel system with 8 cores and 16 Gigabytes of RAM under Linux 4.4.0 kernel.  

\vspace{-3mm}

The goal is to compare the performance of SDN-based approach with alternative distributed solutions proposed in~\cite{rfc7432, hao-bess-evpn-df-handshaking-02} in terms of selecting the DF. The following performance metrics are assessed:

\begin{enumerate}
\item Number of BUM packets which are dropped or duplicated in different methods when the new PE is taking the DF responsibility.

\item Packet loss percentage of BUM traffic when link utilization on the multicast tree (the tree that is shown in red) increases.
\end{enumerate}

\subsection{DF Switch-Over}
\begin{figure*}[!tbh]
    \begin{subfigure}{.50\textwidth}
        \centering
        \includegraphics[width=\textwidth]{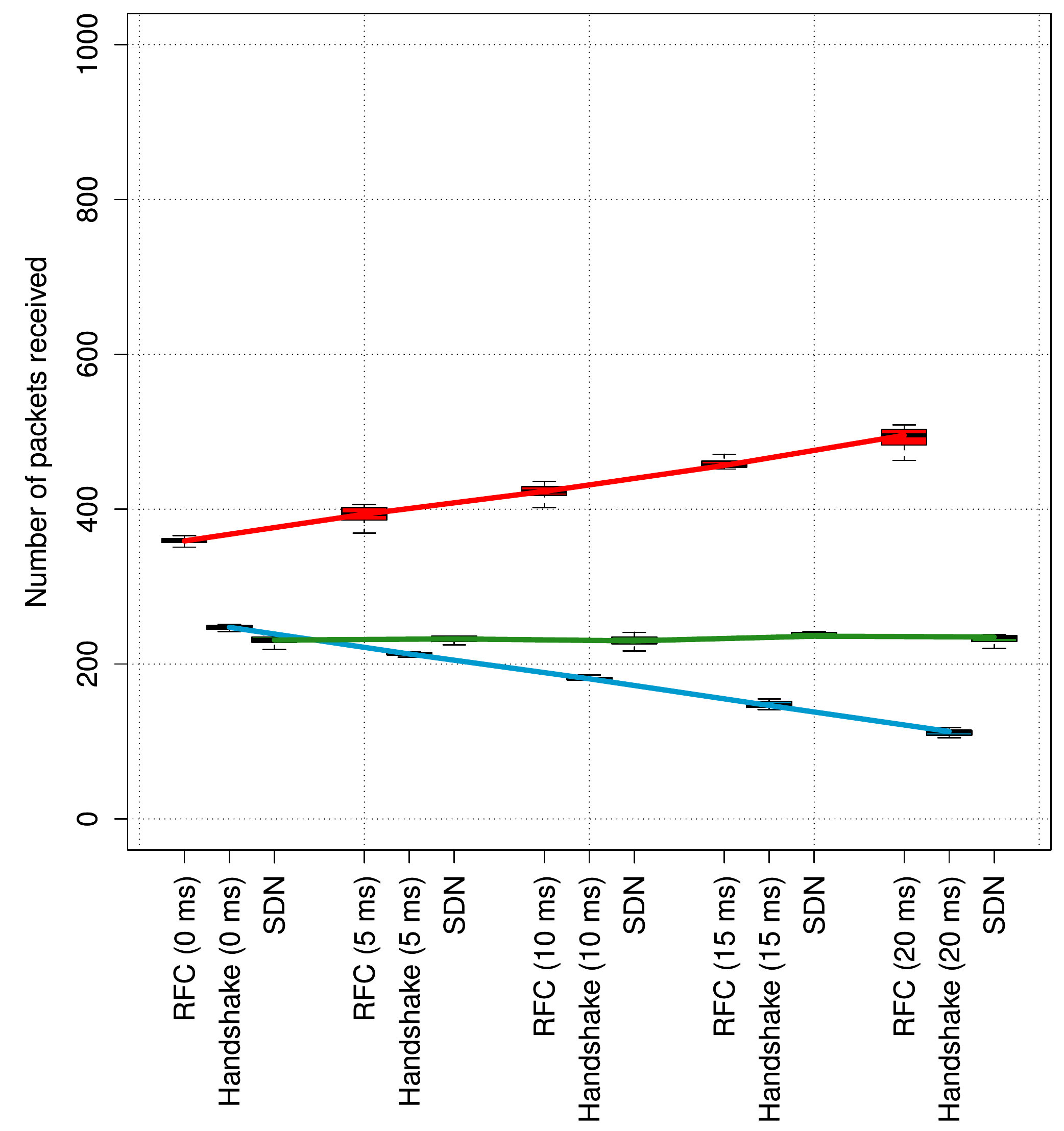}
        \caption{75 Mbps BUM traffic}
        \label{bcube-mptcp}
    \end{subfigure}    
    \hspace{\fill}
    \begin{subfigure}{.50\textwidth}
        \centering
        \includegraphics[width=\textwidth]{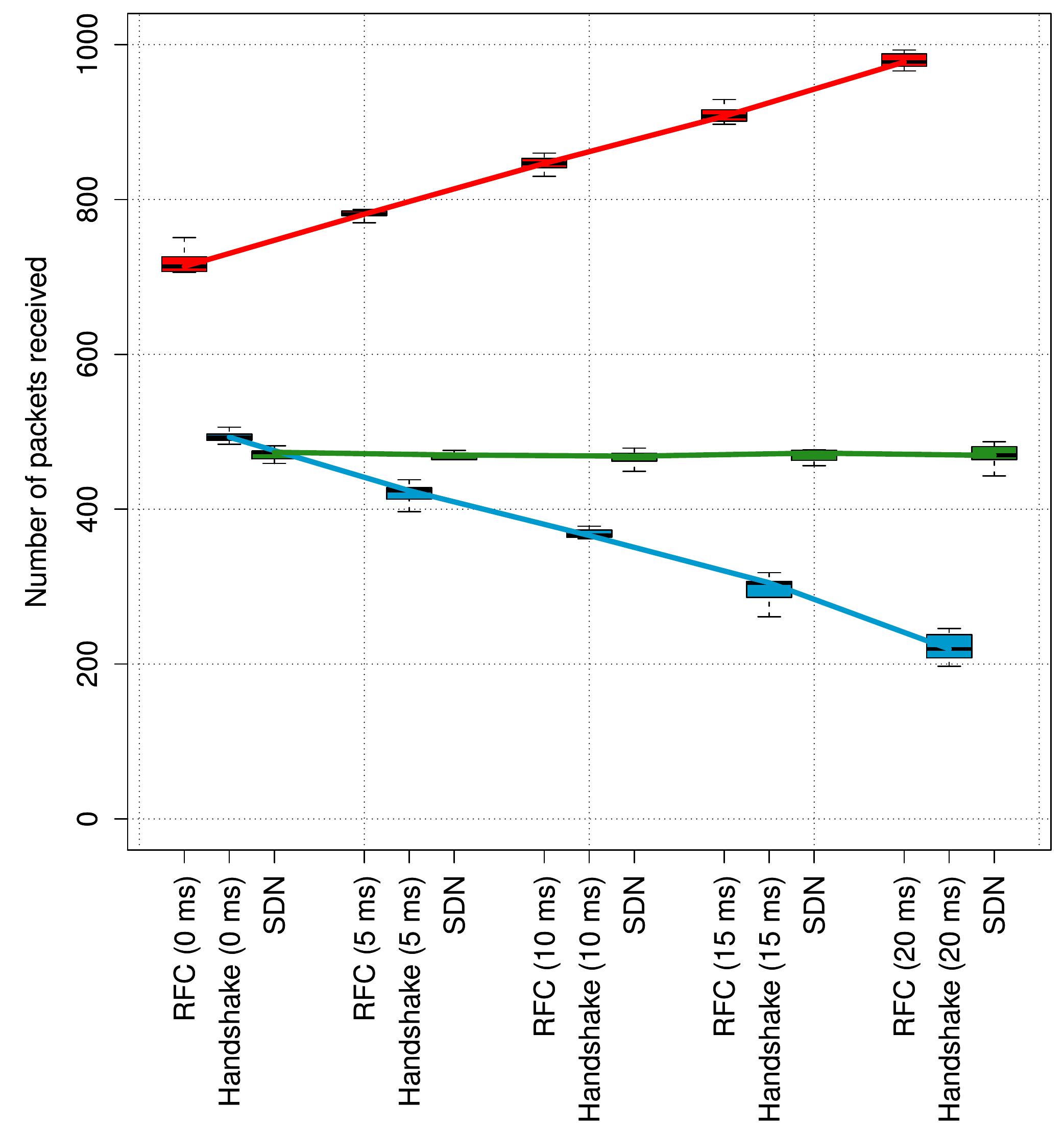}
        \caption{150 Mbps BUM traffic}
        \label{fattree-mptcp}
    \end{subfigure}
    \caption{\label{fig:workload-mobility}Selecting a new DF}
    \vspace{-3mm}
\end{figure*}

As mentioned earlier, the tree with red branches depicted in Figure~\ref{fig:testbed} represents the initial multicast tree that the multicast application considered.  Therefore, PE-1 is the initial DF. The BUM sender starts sending BUM traffic using Iperf to the sink node at two traffic intensities, 75 and 150 Mbps. Then, the new PE (PE-2) is inserted into the DC and the EVIs are configured on that. Now, according to service-carving algorithm or controller decision, we assume that the DF responsibility has to be moved from PE-1 to PE-2. The performance of three approaches are emulated and compared while the delay between PEs is increased from 0 to 20 ms with 5 ms increase at each step: (1) SDN-based solution, (2) service-carving algorithm as proposed in EVPN RFC~\cite{rfc7432}, and (3) handshaking mechanism proposed in~\cite{hao-bess-evpn-df-handshaking-02}. The experiment is conducted 10 times for each BUM traffic intensity, without background traffic, and each emulation lasts for 50 milliseconds. The results are depicted in Figure~\ref{fig:workload-mobility}. The colored lines connect the medians of the box plots.

As shown in the Figure~\ref{fig:workload-mobility}, using the service-carving algorithm the total number of packets captured by the sink node is more than the total number of the packets that we expect\footnote{Approximately 300 packets at 75 Mbps and 600 packets at 150 Mbps.} which implies that there are duplicated packets, because more than one DF was active for a specific time period. In contrast, the sink node received fewer packets than the total sent packets for the handshake proposal and the SDN method. Consequently, there was no DF active for some time interval for both methods. Although all approaches are not ideal, missing BUM packets is more preferred than receiving the same traffic multiple times. Dropping BUM packet may delay establishing the connection and requires the sender to resend the same packets. However, importing BUM packets through multiple routers into the DC is more destructive and leads to an undesirable flooding, overhead in data and control plane (since it may cause MAC flip-flop~\cite{hao-bess-evpn-df-handshaking-02}), and delaying the connection.

By increasing the delay among PE routers from 0 to 20 ms, the RFC and handshake approach lead to more duplicated or lost packets while the SDN approach shows a consistent performance. The reason is that the SDN controller has an out-of-band control connection to the PEs and is not affected by network congestion in the data plane.


\subsection{SDN Controller Triggered DF Change}
One of the major advantages of selecting the DF from the SDN controller in comparison to the proposed distributed solution is that the controller may change the DF dynamically by considering the underlying network status. For instance, the controller can be notified (or infer from statistics gathering) that the downlink capacity from a given PE to the CE is decreasing or that the PE router is becoming over-utilized, which may lead to increased packet loss probability. In such case, the controller may smoothly and dynamically move the DF workload to the PE which is better suited (has more available capacity). On the contrary, in the legacy approach, an elected DF would always send BUM packets to a pre-established multicast tree regardless of PE utilization or multicast tree congestion level.

Akin to the previous experiment, PE-1 is set as the DF for multiple broadcast domains. Source-1 starts sending TCP flow at 850 Mbps to the sink node using Iperf. The purpose is to congest the network links by 85\%. The traffic passes nodes PE-1, CE-1 and TOR-1 to reach the sink node. After 5 seconds, the Source-2 starts sending additional TCP traffic to the sink node through the path TOR-2, CE-1, and TOR-1 in order to further increase the congestion level between CE-1 and TOR-1. Source-2 increases the traffic intensity from 0 to 150 Mbps (25 Mbps increase at each step). Therefore, the congestion level on the link between CE-1 and TOR-1 increases from 85\% to 100\%. Immediately after Source-2, the BUM sender starts sending BUM traffic at two traffic intensities, 50 and 100 Mbps respectively.

The multicast application calculates multicast tree weights every 5 seconds. Besides the total tree weight, the multicast application considers the link congestion level in the tree to select the active tree. When the red tree weight is not smaller than the alternative tree (green) and the congestion level for one or some of its links is above a threshold (for this experiment 95\%), the multicast application changes the active multicast tree. Consequently, the L2VPN Service changes the DF.

\begin{figure}[tbh]
\begin{centering}
\includegraphics[width=0.43\textwidth]{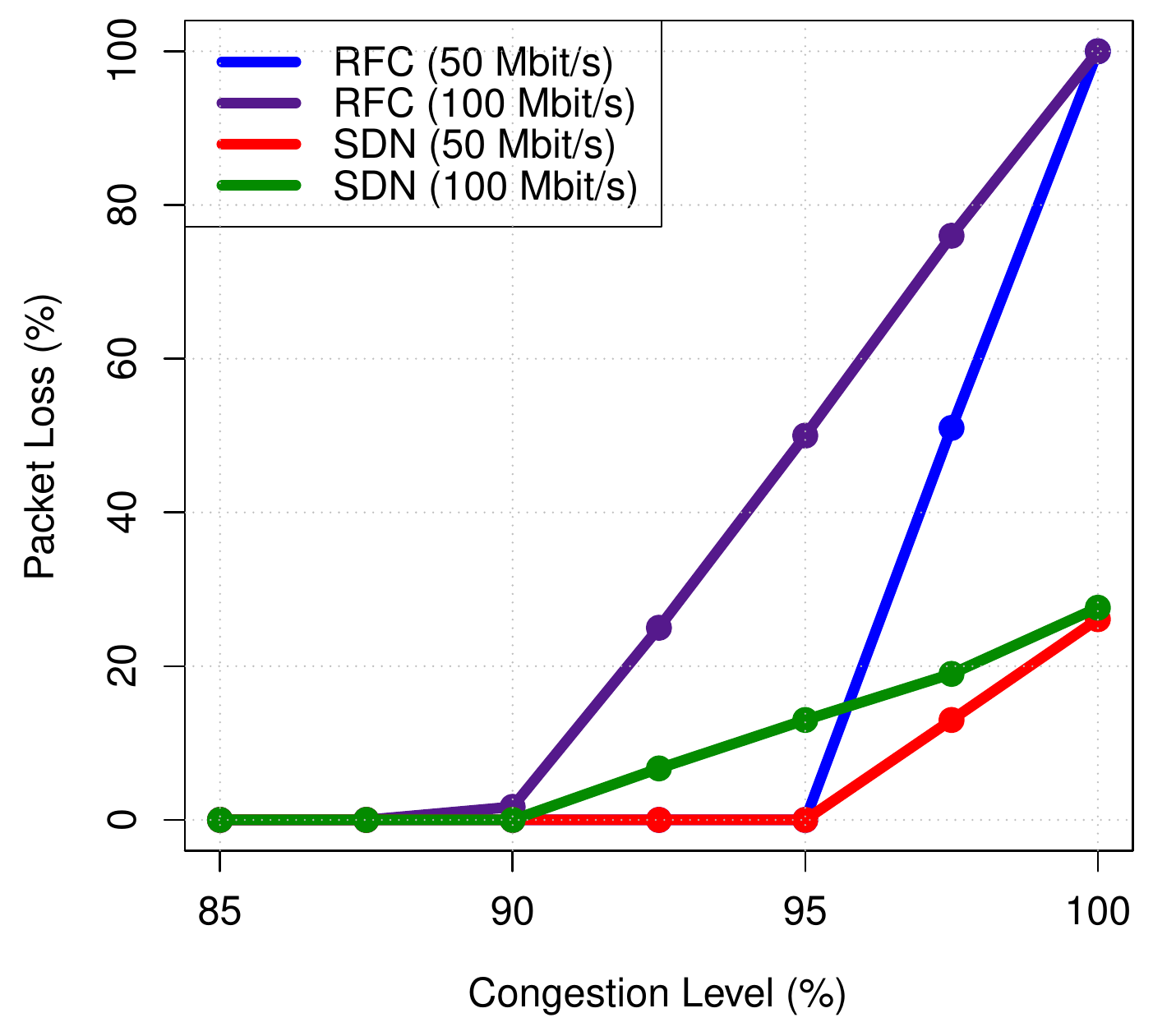}
\par\end{centering}
\caption{\label{fig:change-df}Change DF according to multicast tree weight}
\vspace{-3mm}
\end{figure}

As shown in Figure~\ref{fig:change-df}, the SDN approach dynamically moves the DF responsibility from PE-1 to PE-2, when the multicast application finds the multicast tree with lower weight. In contrast, the RFC approach continues leveraging the old multicast tree. Although SDN approach also causes some packet loss, the packet loss percentage is approximately 70\% higher when the DF responsibility is not moved in the RFC approach. By saving more BUM packet from being dropped the SDN approach helps considerably DC providers by removing the need for retriggering the same traffic again. Moreover, the application performance would be enhanced and the initial delay to establish a connection would be decreased.

\section{Conclusions}
\label{sec:conclusion}
Although Ethernet VPN has addressed new data center interconnection requirements, it introduces new challenges which have to be resolved. All active redundancy mode enables load balancing of layer 2 unicast traffic across all the multi-homed links on and toward a customer edge device. However, it would be highly undesirable for all provider edge routers to forward multi-destination traffic (e.g., broadcast) and so only one, which is known as designated forwarder, is responsible to do so in order to prevent traffic duplication. Proposed solutions for selecting the DF are based on running a distributed algorithm on provider edges which has a number of problems. This paper investigated how Software-Defined Networking (SDN) architecture could enhance the designated forwarder selection procedure and mitigate the corresponding problems. Our experiments indicated that the SDN approach can reduce packet loss of multi-destination traffic. Moreover, the SDN-based solution shows consistent performance and would not be affected by network delay and message passing among edge routers. Finally, we have shown that SDN-based solution could change designated forwarder deterministically or dynamically according to underlying network status.

\section*{ACKNOWLEDGMENT}
This research has been funded by the Knowledge Foundation of Sweden through the project HITS.

{\small{}\bibliographystyle{./IEEEtran}
\bibliography{refs}
}{\small \par}

\end{document}